\documentclass[9pt,twocolumn,twoside]{opticajnl}

\journal{opticajournal} 

\setboolean{shortarticle}{true}

\title{Photonic random walks with traps}

\author[1,2,*]{Stefano Longhi}

\affil[1]{Dipartimento di Fisica, Politecnico di Milano, Piazza L. da Vinci 32, I-20133 Milano, Italy}
\affil[2]{IFISC (UIB-CSIC), Instituto de Fisica Interdisciplinar y Sistemas Complejos - Palma de Mallorca, Spain}

\affil[*]{stefano.longhi@polimi.it}

\begin{abstract}
Random walks behave very differently for classical and quantum particles. Here we unveil a ubiquitous distinctive behavior of random walks of a  photon in a one-dimensional lattice in the presence of a finite number of traps, at which the photon can be destroyed and the walk terminates. While for a classical random walk the photon is unavoidably destroyed by the traps, for a quantum walk the photon can remain alive and the walk continues forever. Such an intriguing behavior is illustrated by considering photonic random walks in  synthetic mesh lattices with controllable decoherence, which enables to switch from quantum to classical random walks.
\end{abstract}

\setboolean{displaycopyright}{false} 

\begin{document}

\maketitle

{\em Introduction.}  Random walks (RW) are essential tools in modeling various phenomena in physics, biology, and computer science \cite{R1,R2,R3,R4}.
Most of these
applications use classical random walks (CRW), in which
quantum mechanical principles are not considered.
Quantum random walks (QRW) extend the concept of CRW into the quantum realm \cite{R5,R6,R7,R8,R9,R10,R11}, allowing the particle to be in multiple states simultaneously. 
QRW exhibit behaviors that differ significantly from CRW, including interference effects and faster spreading of the particle probability distribution, which 
could be of relevance in diverse quantum applications, such as quantum search algorithms and quantum state engineering.
To this regard, photonics has provided over the past decade a fascinating platform for implementing QRW \cite{R12,R13,R14,R15,R16,R17,R18,R19,R20,R21,R22,R23,R24,R25}, offering the possibility of controlling decoherence and transitioning from QRW to CRW \cite{R16,R17}.\par
Random walks with traps \cite{R2,R26,R27,R28,R29,R30,R31,R32,R33,R34,R35,R36,R37} are a variant of the RW problem where certain sites are designated as traps. In this case a particle starts at a certain initial site and moves randomly to adjacent sites at each time step. However, when the particle reaches a trap site, it is annihilated and the walk terminates.  Random walks with traps find important applications in understanding a wide variety of phenomena, such as diffusion processes in physics, modeling the spread of diseases in epidemiology, analyzing financial markets, and studying the behavior of molecules in biology, to mention a few. The different behaviors of CRW versus QRW with traps have received little attention so far. Previous works \cite{R32,R33,R34,R35,R37} mainly highlighted the decoherence role played by traps in QRW \cite{R33,R34}, the tendency of QRW to avoid traps owing to interference \cite{R35}, and some topological properties of non-Hermitian QRW in bipartite lattices \cite{R37}. Analytical results have been given in very special configurations \cite{R32,R35,R37}. The problem of the fate of the walker in infinite or semi-infinite lattices with a finite number of irregularly-displaced traps  remains so far unsolved. \par
In this Letter we consider RW of photons on a one-dimensional lattice with traps and show that, for an arbitrary finite number of irregularly-displaced traps, in a CRW the photon is unavoidably destroyed by the traps, while in a QRW the photon can survive and the walk continues forever. Such an intriguing behavior is illustrated by considering photonic RW in  experimentally-accessible synthetic mesh lattices with controllable decoherence, allowing to switch from QRW to CRW.

{\em Classical versus quantum walks with traps: model and main results.}
 Let us consider a continuous-time RW of a particle, such as a photon, on a one-dimensional lattice with a {\em finite number} of  {\em irregularly-spaced} traps on the lattice, which can annihilate the particle at some given rates  [Fig.1(a)]. For the sake of definiteness, we consider an infinite lattice, however the results can be extended {\em mutatis mutandis} to the case of a semi-infinite lattice. At initial time $t=0$, let us assume that the particle is at the site $n=n_0$ of the lattice, and let $P(t)$ be the survival probability of the particle at subsequent times $t$. The main question is whether there is a chance for the particle to remain alive and continue the walk forever, i.e. whether $P_{\infty} \equiv P( t \rightarrow \infty)$ is non-vanishing.  The answer to this question is set by the classical or quantum nature of the walker, i.e. whether the particle can or cannot be in a superposition state  [Fig.1(a)].\\
For a classical RW, the probability $p_n(t)$ for the particle to be at site $n$ at time $t$ is governed by the classical master equation \cite{R1,R2,R3,R35}
\begin{equation}
\frac{dp_n}{dt}=J(p_{n+1}+p_{n-1})-2Jp_n-\gamma_n p_n
\end{equation}
 \begin{figure}[h]
 \centering
    \includegraphics[width=0.42\textwidth]{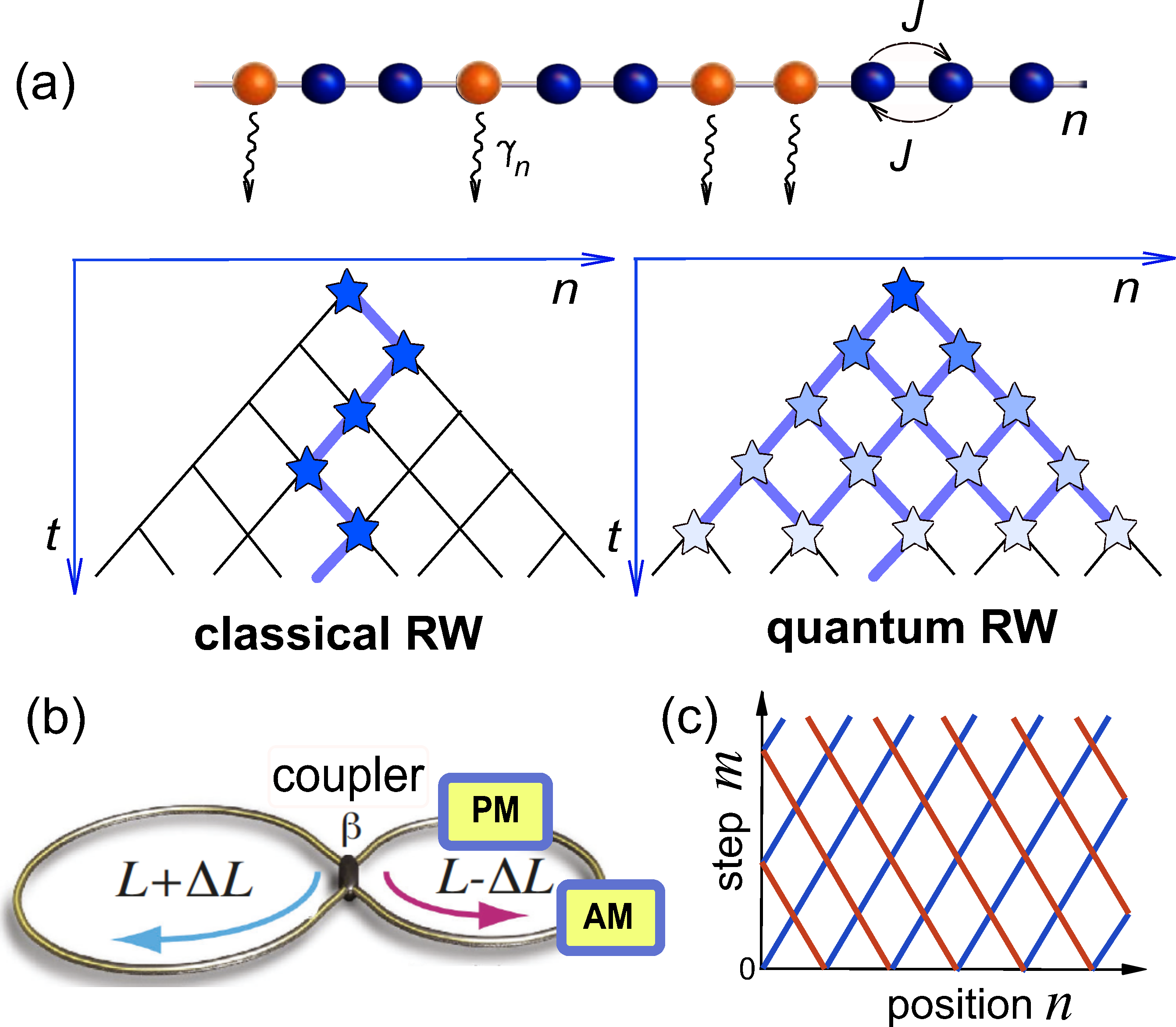}
   \caption{ \small (a) Schematic of a random walk for a classical or quantum particle (indicated by a star in the cartoon) on a one-dimensional lattice with traps. The hopping rate between adjacent sites is $J$. The traps are indicated by the filled red circles. $\gamma_n$ are the annihilation rates of the particle at the trap sites. (b) Schematic of the coupled fiber
loops that realize photonic RW with traps
in the synthetic dimension. $\beta$ is the coupling angle between the fiber loops,
AM and PM are amplitude and phase modulators. The AM controls the loss rates at the trap sites, whereas the PM  is used to switch from QRW to CRW by the introduction of dephasing effects in the coherent wave dynamics. (c)  Schematic of the synthetic mesh lattice.}
\end{figure}
where $J$ is the hopping rate between adjacent sites and $\gamma_n$ is the annihilation rate at site $n$ of the lattice. The survival probability of the particle at time $t$ is given by $P(t)=\sum_{n=-\infty}^{\infty} p_n(t)$. Clearly, in the absence of the traps, i.e. for $\gamma_n=0$, one has $P(t)=1$ at any time $t$ and the initial excitation, $p_n(t=0)=\delta_{n,n_0}$, spreads diffusively in the lattice and for large $t$ the probability distribution $p_n(t)$ has the Gaussian profile $p_n(t) \simeq (4 \pi Jt )^{-1/2} \exp[-(n-n_0)^2/(4Jt)]$ characteristic of a diffusive motion. When we have a finite number of traps on the lattice, a central result of CRW is that the survival probability $P_{\infty}$ always vanishes, for any {\em arbitrary} setting of the traps and annihilation rates $\gamma_n$, with a universal asymptotic algebraic decay $P(t) \sim 1/ \sqrt{Jt} $ at long times. This means that for a CRW the particle does not have any chance to survive and to continue the walk forever. The proof thereof, which is given in the Supplemental document, is rather lengthy and requires considering the asymptotic behavior of the solution $p_n(t)$ to Eq.(1) using spectral methods.
As illustrative examples, Fig.2(b) depicts the numerically-computed evolution of the survival probability $P(t)$ in a lattice with four traps in a few different arrangements, illustrated in Fig.2(a). The figure clearly indicates an irreversible decay of $P(t)$ toward zero with an asymptotic behavior $P(t) \sim 1/ \sqrt{Jt}$ at long times, independent of the trap configuration.\\
A very different scenario is found for a QRW.  The continuous-time RW of a quantum particle 
is governed by the effective non-Hermitian Schr\"odinger equation \cite{R35,R37}
\begin{equation}
i \frac{d \psi_n}{dt}=J(\psi_{n+1}+\psi_{n-1})-i \frac{\gamma_n}{2} \psi_n \equiv H \psi_n \label{Schro}
\end{equation}
for the amplitude probabilities $\psi_n(t)$ to find the particle at site $n$ at time $t$, where $J$ is the hopping rate between adjacent sites, $\gamma_n \geq 0$ is the loss (annihilation) rate at site $n$, and $H=H_{n,m}=J(\delta_{n,m+1}+\delta_{n,m-1})-i \gamma_n \delta_{n,m}$ is the non-Hermitian matrix Hamiltonian of the lattice. In this case, it can be demonstrated in a very general way that, contrary to a CRW, the survival probability $P(t)=\sum_{n=-\infty}^{\infty} | \psi_n(t)|^2$ does not vanish as $ t \rightarrow \infty$, i.e. the quantum particle always has a chance to survive and to walk for the ever on the lattice. The proof of such a theorem entails an asymptotic analysis of the solution to Eq.(\ref{Schro}) as $t \rightarrow \infty$ using the steepest descent method, and it is given in the Supplemental document. Figure 2(c) depicts, as an illustrative example, the numerically-computed evolution of the survival probability $P(t)$ for a QRW in the trap arrangements  of Fig.2(a). The figure clearly indicates that, contrary to the CRW, in the QRW the survival probability $P(t)$ settles down to a finite and non-vanishing value as $t \rightarrow \infty$. From a physical viewpoint, the different result for QRW vs CRW stems from the fact that in the former case the spreading in the lattice is ballistic with a finite speed, and once the particle has escaped from the trap region (which includes a finite number of traps), it can quickly move away from the traps indefinitely. Conversely, in a CRW with unbiased coin the spreading is diffusive and the walker repeatedly comes back to the traps,  even after passing through them, finally being destroyed. It should be mentioned that a non-vanishing survival probability is observed in a CRW when there is a bias in the coin toss \cite{R31}: in this case the walker drifts along the lattice and,  once it has escaped from the trap region, it can quickly move away from the traps indefinitely. 
 \begin{figure}[htbp]
 \centering
    \includegraphics[width=0.45\textwidth]{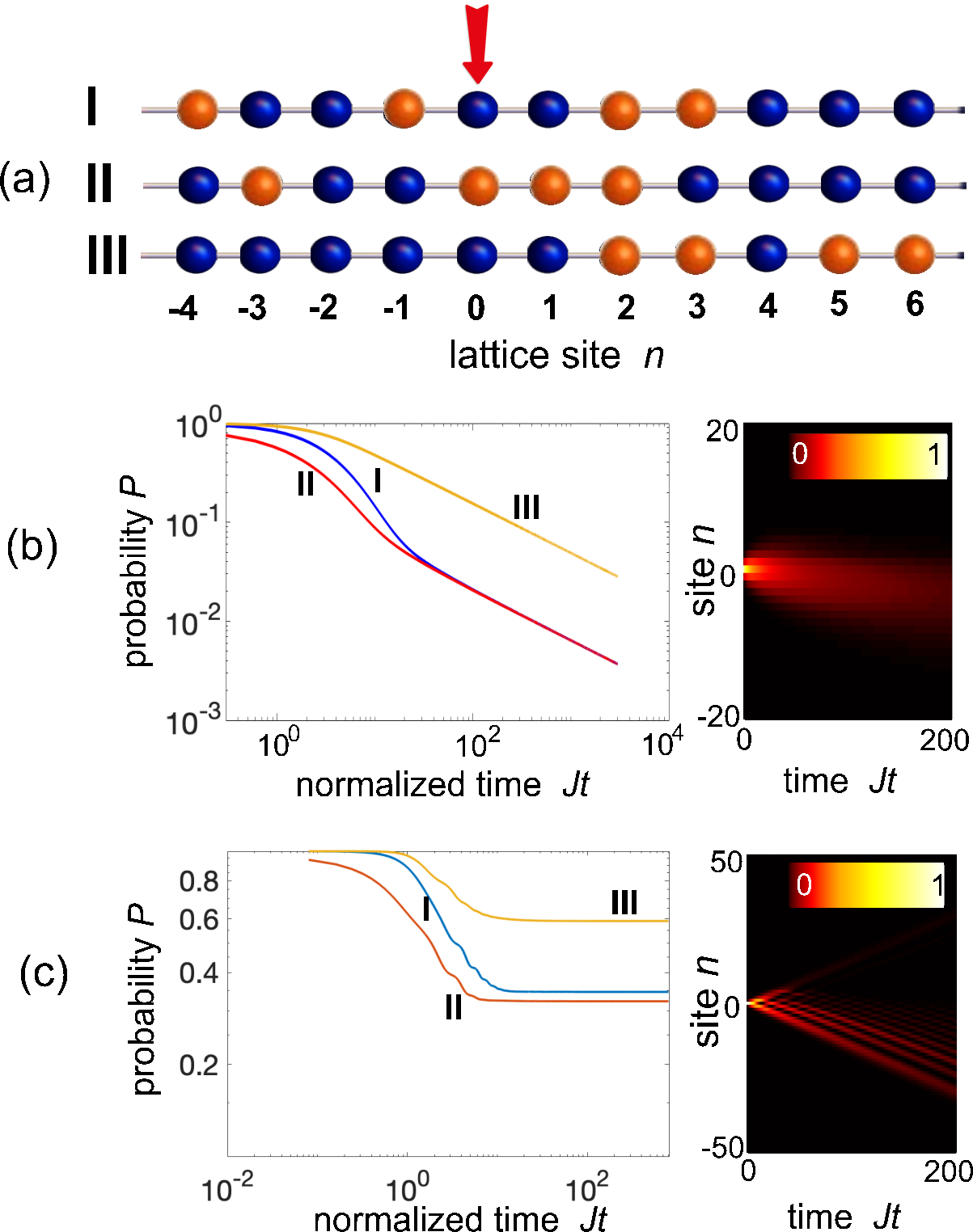}
   \caption{  \small (a) Schematic of three lattices (I,II,III) with four traps at different locations, indicated by the red filled circles. The loss rates in the four traps, for increasing values of lattice site index $n$, are $\gamma_n/J=1, 0.4, 1.5$ and 0.6. The particle starts the walk at the lattice site $n=0$, indicated by the vertical arrow. (b)  Left panel: Temporal evolution of the survival probability $P(t)$ on a log scale for a CRW in the three lattice configurations I, II and III. Right-panel: snapshot of the temporal evolution of site occupation probabilities $p_n(t)$ on a pseudocolor map for the trap configuration III. Note that  spreading is diffusive and the walker is not able asymptotically to avoid the trap region, resulting in an irreversible decay of the survival probability. (c) Same as (b) but for a QRW. Note that in this case the spreading is ballistic and the walker is able to drift far apart from the trap region, resulting in a non-vanishing probability to survive and to continue the walk forever. }
\end{figure}

{\em Photonic implementation of random walks with traps.}
An experimentally-accessible platform to demonstrate the different fate of classical vs quantum particles undergoing a RW with traps is provided by photonic RW in synthetic mesh lattices 
  based on light pulse dynamics in coupled fiber loops \cite{R38,R39,R40,R41}, where transition from quantum to classical walks can be controlled by the introduction of random dynamic phase changes \cite{R17}. 
  We remark that the RW of a single photon in such a system is exactly reproduced by using classical optical fields, so that we may consider light dynamics using a mean-field (classical) model.
  The system  consists of two fiber loops of slightly different lengths that are connected by a fiber coupler with a coupling angle $\beta$ [Fig.1(b)]. Phase and amplitude modulators are placed in one of the two loops, providing the desired control of the phase and amplitude of the traveling pulses.  When a single pulse is injected from one loop, it will evolve into a pulse train after successive pulse splitting at central coupler, circulating in two loops and interference at the coupler again. For two loops with lengths $L \pm \Delta L$, the pulse physical time is discretized as $t_n^m= mT +n\Delta T$, where $T= L/c$ is mean travel time and $\Delta T=\Delta L/c \ll T$ is travel-time difference in two loops. The pulse dynamics can thus be mapped into a "link-node" lattice model $(n, m)$ as shown in Fig.1(c), where $n$, $m$ denote the transverse lattice site and longitudinal evolution step. The leftward/rightward links towards the node correspond to pulse circulations in short/long loops and scattering at each node corresponds to pulse interference at the coupler. Light dynamics is described by the set of discrete-time equations \cite{R38,R39,R40,R41}
 \begin{eqnarray}
 u^{(m+1)}_n & = & \left(   \cos (\beta) u^{(m)}_{n+1}+i \sin (\beta) v^{(m)}_{n+1}  \right)  \exp (-i\phi_{n}^{(m)}- \gamma_n ) \;\;\;\;\;\;  \\
 v^{(m+1)}_n & = & \left(   \cos (\beta) v^{(m)}_{n-1}+i \sin (\beta) u^{(m)}_{n-1}  \right)
 \end{eqnarray}
 where $u_n^{(m)}$ and $v_n^{(m)}$ are the pulse amplitudes at discrete time step $m$ and lattice site $n$ in the two fiber loops, $2 \phi_n^{(m)}$ is the phase change impressed by the phase modulator at site $n$ and time step $m$, and $ \gamma_n \geq 0 $ is the loss rate at lattice site $n$ impressed by the amplitude modulator, which is assumed independent of time step $m$. The sites with non-vanishing $\gamma_n$ correspond to the trap sites. The phase modulator is used to introduce decoherence (dephasing) in the wave propagation, and thus to switch from a quantum (coherent) to classical (incoherent) walk \cite{R17,R42}.  The QRW is obtained by letting $\phi_n^{(m)}=0$, i.e. switching off the phase modulator. In this case, assuming a coupling angle $\beta$ close to $\pi /2$ and $\gamma_n \ll 1$, the discrete-time QRW of light pulses described by Eqs.(3) and (4) reproduces the continuous-time QRW, defined by Eq.(2), with an hopping rate $ \pm J=\pm (1/2) \cos \beta$ and loss rates $\gamma_n$. Specifically, after elimination of the amplitudes $v_n^{(m)}$ from Eqs.(3) and (4), the solution to the pulse amplitudes $u_n^{(m)}$ can be written as 
 \begin{equation}
 u_{n}^{(m)} \simeq (i)^m \left\{  \psi_{n}^{(+)}(m)+(-1)^m \psi_n^{(-)}(m) \right\}, 
 \end{equation}
 where the envelopes $\psi_n^{(\pm)}(m)$ vary slowly with time step $m$ and satisfy the decoupled Schr\"odinger-like wave equations (see \cite{R42} and Supplemental document for technical details)
 \begin{equation}
 i \frac{d \psi_{n}^{(\pm)}}{dm}= \pm J \left( \psi_{n+1}^{(\pm)}+\psi_{n-1}^{(\pm)} \right) - i \frac{\gamma_n}{2} \psi_{n}^{(\pm)}
 \end{equation}
which is precisely the continuous-time QRW with traps as given by Eq.(2).\\
The photonic CRW is obtained by switching on the phase modulator and assuming  $\varphi_n^{(m)}$  uncorrelated stochastic phases with uniform distribution in the range $(-\pi, \pi)$. Such stochastic phases impressed at each time step in the light dynamics introduce dephasing effects, spoiling out phase coherence and interference effects in the dynamics \cite{R17,R42}.   
Under incoherent dynamics, light evolution is described by the following map \cite{R42}
 \begin{eqnarray}
X_n^{(m+1)} & = & \left(  X_{n+1}^{(m)} \cos^2 \beta + Y_{n}^{(m)}  \sin^2 \beta  \right) \exp(-2 \gamma_n) \\
Y_n^{(m+1)} & = & X_{n}^{(m)}\sin^2 \beta  + Y_{n-1}^{(m)} \cos^2 \beta  
\end{eqnarray}
  for the light pulse intensities $X_n^{(m)}=\overline{|u_n^{(m)}|^2}$ and $Y_n^{(m)}=\overline{|v_{n+1}^{(m)}|^2}$ in the two fiber loops, where the overline denotes statistical average. Assuming as in the QRW case a coupling angle $\beta$ close to $\pi/2$ and small losses $\gamma_n \ll 1$, the occupation probability of the photon at site $n$ of the synthetic lattice and at discrete time step $m$, in either one of the two fiber loops, i.e. $p_n^{(m)}=X_n^{(m)}+Y_n^{(m)}$, varies slowly with $m$ and satisfies the master equation (see Supplemental document)
\begin{equation}
\frac{d p_n}{dt}=J(p_{n+1}+p_{n-1})-(2J+\gamma_n) p_n
\end{equation}  
with a hopping rate $J=(1/2) \cos^2 \beta$. This is precisely the master equation (1) of a continuous-time CRW on a line with traps described by Eq.(1). The survival probability of the photon undergoing the CRW is simply given by $P(m)=\sum_n p_n(m)$, with the time $t$ is replaced by the discrete  time step $m$ .
 \begin{figure}[h]
  \centering
    \includegraphics[width=0.49\textwidth]{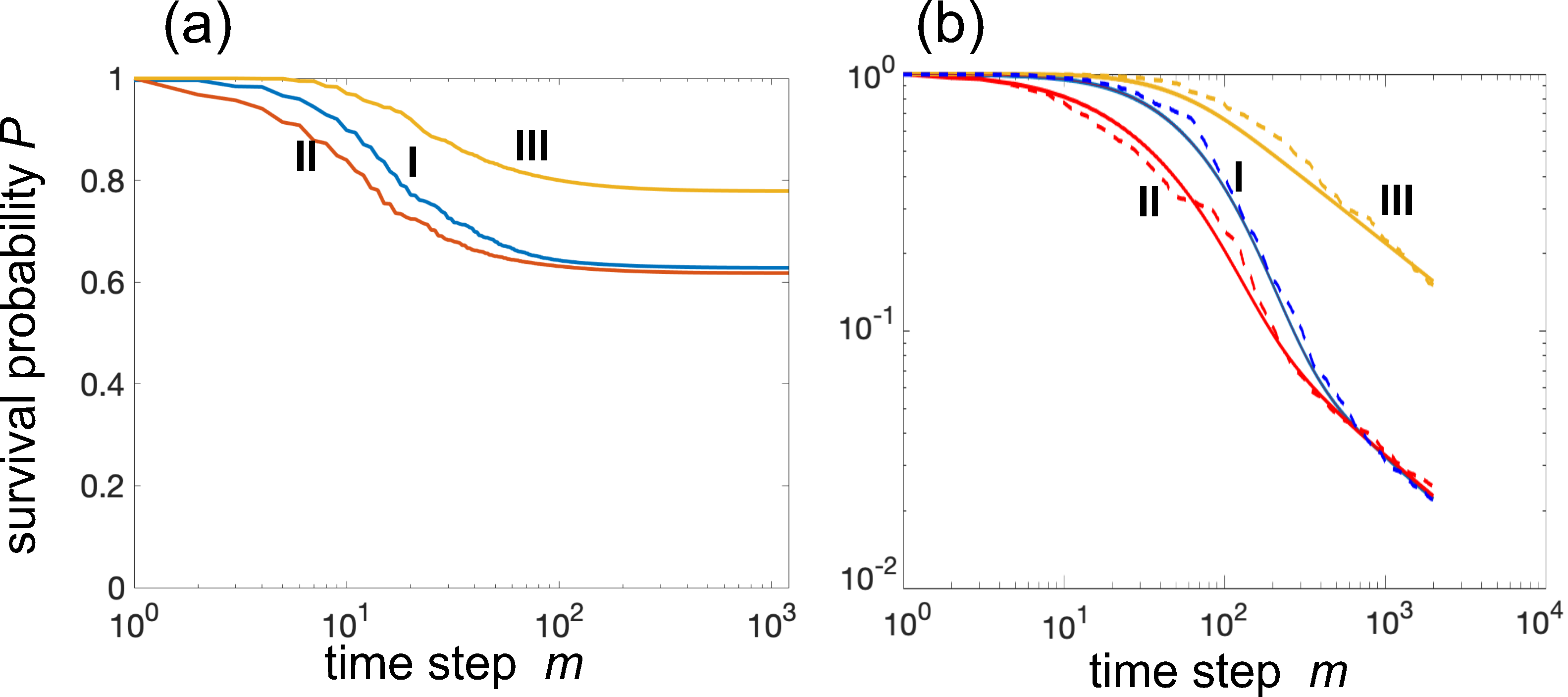}
    \caption{\small Behavior of the survival probability $P(m)$ in a photonic RW on a synthetic mesh lattice, realized in the coupled fiber loop setup schematically shown in Fig.1(b). Coupling angle $\beta= 0.8 \times \pi/2$. The loss rates $\gamma_n$ impressed by the AM modulator are applied at four sites of the synthetic lattice, in the three configurations I, II, and III as in Fig.2(a). The values of $\gamma_n$ are (for increasing values of site index $n$) $\gamma_n=0.1, 0.04, 0.15$ and 0.06.
(a) Behavior of the survival probability $P(m)$ when the phase modulator PM is switched off, corresponding to QRW. (b) Same as (a), but when the phase modulator is switched on, phase randomization is applied at every time step, and the RW becomes classical. The solid curves are obtained after statistical average over 1000 stochastic realization of disorder, and correspond to the behavior predicted by the master equations (7) and (8). The dashed lines depict the behaviors of $P(m)$ for single realizations of randomized phases.}
    \end{figure}
 The theoretical predictions have been confirmed by full numerical simulations of Eqs.(3) and (4), i.e. beyond the continuous-time limit of the RW. Typical numerical results, showing the behavior of the survival probability in quantum and classical RW of light pulses in the fiber loop setup, are shown in Fig.3. Like in Fig.2(a), we introduce in the lattice four traps in three different configurations I, II and III. Single pulse excitation is assumed by taking $u_n^{(0)}=\delta_{n,0}$ and $v_n^{(0)}=0$.
 The largest propagation step assumed in the simulations is $m=2000$, which is experimentally feasible keeping the required coherence over the entire propagation window \cite{R40}.
  When the phase modulator is switched off, the photon undergoes a coherent walk on the lattice (QRW). The survival probability does not decay and asymptotically settles down to a finite and constant value  [Fig.3(a)], as a result of the ballistic spreading of the photon far from the traps. Conversely, when the phase modulator is switched on and stochastic phases are applied at each time step $m$, the behavior of the survival probability, averaged over many different realizations of stochastic phases, irreversibly decays toward zero, as shown in Fig.3(b). In the figure, the solid lines depict the decay behavior of the survival probability, 
  \begin{equation}
  P(m)= \overline{\sum_n (|u_{n}^{(m)}|^2+|v_{n+1}^{(m)}|^2 }), 
  \end{equation}
   taking a statistical average over 1000 realizations of stochastic phases, reproducing the decay dynamics as described by the classical master equations (7) and (8).  The dashed curves in the figure show the corresponding behavior of the survival probability $P(m)= \sum_n (|u_{n}^{(m)}|^2+|v_{n+1}^{(m)}|^2 )$ for a single trajectory, i.e. a single realization of random phases.

\noindent

{\em Conclusion.}
 In conclusion, we  unravelled a universal distinct behavior of classical versus quantum photonic random walks in one-dimensional lattices in the presence of a finite number of traps. While in the classical RW the photon is unavoidably annihilated by the traps and the walk terminates, in the quantum RW the photon has a chance to survive and to walk forever on the lattice. Such a property has been demonstrated to hold for arbitrary arrangements of the traps and physically explained by the inability of the particle in a classical RW to asymptotically depart from the trap region, finally resulting in an unavoidable particle destruction. The predicted universal phenomenon provides a major distinctive feature of classical vs quantum walks, beyond the well known different spreading  laws (diffusive versus ballistic), and should be experimentally observable in photonics using synthetic lattices. The present results could motivate further theoretical and experimental directions. For  example, it would be interesting to investigate the role of quantum interference and particle statistics \cite{R18} in photonic RW with traps, as well as to explore with photons intermittency phenomena in randomly-distributed traps \cite{R43}.\\
\\
\noindent
{\bf Disclosures}. The author declares no conflicts of interest.\\
\\
{\bf Data availability}. No data were generated or analyzed in the presented research.\\
\\
{\bf Funding}. Agencia Estatal de Investigacion (MDM-2017-0711).\\
\\
{\bf Supplemental document}. See Supplement 1 for supporting content.

\newpage


 {\bf References with full titles}\\
 \\
 \noindent
1. G.F. Lawler and V. Limic, {\it Random Walk: A Modern Introduction} (Cambridge University Press, Cambridge, UK, 2012).\\
2. J.W. Haus and K.W. Kehr, Diffusion in regular and disordered lattices, Phys. Rep. {\bf 150}, 263 (1987).\\
3. N. Masuda, M.A. Porter, and R.Lambiotte, Random walks and diffusion on networks, Phys. Rep. {\bf 716}, 1 (2017). \\
4. E.A. Codling, M.J. Plank, and S. Benhamou, Random walk models in biology, R. Soc. Interfac. {\bf 5}, 813 (2008).\\
5. Y. Aharonov, L. Davidovich, and N. Zagury,  
 Quantum random walks,
Phys. Rev. A {\bf 48}, 1687 (1993).\\
6. A.M. Childs, E. Farhi, and S. Gutmann, 
An example of the difference between quantum and classical random walks, Quantum Inf. Process. {\bf 1}, 35 (2002).\\
7. J. Kempe, Quantum random walks: An introductory overview, Contemporary Phys. {\bf 44}, 307 (2003).\\
8. N. Shenvi, J. Kempe, and K. Birgitta Whaley, Quantum random-walk search algorithm,
Phys. Rev. A {\bf 67}, 052307 (2003).\\
9. A.M. Childs, Universal Computation by Quantum Walk,
Phys. Rev. Lett. {\bf 102}, 180501 (2009).\\
10. Oliver M\"ulken and A. Blumen,
Continuous-time quantum walks: Models for coherent transport on complex networks, Phys. Rep. {\bf 502}, 37 (2011). \\
11. S.E. Venegas-Andraca, Quantum walks: a comprehensive review, Quantum Inf. Process {\bf 11}, 1015 (2012).\\
12. H.B. Perets, Y. Lahini, F. Pozzi, M. Sorel, R. Morandotti, and Y. Silberberg, 
Realization of Quantum Walks with Negligible Decoherence in Waveguide Lattices,
Phys. Rev. Lett. {\bf 100}, 170506 (2008).\\
13. A. Peruzzo, M. Lobino, J.C.F. Matthews, N. Matsuda, A. Politi, K. Poulios, X.-Q. Zhou, Y. Lahini, N. Ismail, K. W\"orhoff, Y. Bromberg, Y. Silberberg, M.G. Thompson, and J.L. O'Brien,
Quantum walks of correlated particles, Science {\bf  329}, 1500 (2010).\\
14. A. Schreiber, K. N. Cassemiro, V. Potocek, A. Gabris, P. J. Mosley, E. Andersson, I. Jex, and Ch. Silberhorn, Photons Walking the Line: A Quantum Walk with Adjustable Coin Operations,
Phys. Rev. Lett. {\bf 104}, 050502 (2010).\\
15. A. Schreiber, A. Gabris, P.P. Rohde, K. Laiho, M. Stefanak, V. Potocek, C. Hamilton, I. Jex, and C. Silberhorn, A 2D Quantum Walk Simulation of Two-Particle Dynamics,
Science {\bf 336}, 55 (2012).\\
16. M. A. Broome, A. Fedrizzi, B. P. Lanyon, I. Kassal, A. Aspuru-Guzik, and A. G. White, Discrete Single-Photon Quantum Walks with Tunable Decoherence,
Phys. Rev. Lett. {\bf 104}, 153602 (2010).\\
17. A. Schreiber, K. N. Cassemiro, V. Potocek, A. Gabris, I. Jex, and Ch. Silberhorn, 
Decoherence and Disorder in Quantum Walks: From Ballistic Spread to Localization,
Phys. Rev. Lett. {\bf 106}, 180403 (2011).\\
18. L. Sansoni, F. Sciarrino, G. Vallone, P. Mataloni, A. Crespi, R. Ramponi, and R. Osellame, Two-particle bosonic-fermionic quantum walk via 3D integrated photonics, Phys. Rev. Lett. {\bf 108}, 010502 (2012).\\
19. F. Cardano, F. Massa, H. Qassim, E. Karimi, S. Slussarenko, D. Paparo, C. de Lisio, F. Sciarrino, E. Santamato, R.W. Boyd, and L. Marrucci, Sci. Advances {\bf 1}, e1500087 (2015).\\
20. A. D'Errico, F. Cardano, M. Maffei, A. Dauphin, R. Barboza, C. Esposito, B. Piccirillo, M. Lewenstein, P. Massignan, and L. Marrucci, Two-dimensional topological quantum walks in the momentum space of structured light, Optica {\bf 7}, 108 (2020).\\
21. F. Cardano, A. D'Errico, A. Dauphin, M. Maffei, B. Piccirillo, C. de Lisio, G. De Filippis, V. Cataudella, E. Santamato, L. Marrucci, M. Lewenstein, and P. Massignan, Detection of Zak phases and topological invariants in a chiral quantum walk of twisted photons, Nature Commun. {\bf 8}, 15516 (2017).\\
22. X. Wang, L. Xiao, X. Qiu, K. Wang, W. Yi, and P. Xue, Detecting topological invariants and revealing topological phase transitions in discrete-time photonic quantum walks,
Phys. Rev. A {\bf 98}, 013835 (2018).\\
23. K. Wang, X. Qiu, L. Xiao, X. Zhan, Z. Bian, W. Yi, P. Xue, Simulating dynamic quantum phase transitions in photonic quantum walks,
Phys. Rev. Lett. {\bf 122}, 020501 (2019).\\
24. Z.-Q. Jiao, S. Longhi, X.-W. Wang, J. Gao, W.-H Zhou, Y. Wang, Y.-X. Fu, L. Wang, R.-J. Ren, L.-F. Qiao, and X.-M. Jin, Experimentally Detecting Quantized Zak Phases without Chiral Symmetry in Photonic Lattices,
Phys. Rev. Lett. {\bf 127}, 147401 (2021).\\
25. C. Esposito, M.R. Barros, A. Duran Hernandez, G. Carvacho, F. Di Colandrea, R. Barboza, F. Cardano, N. Spagnolo, L. Marrucci, and F. Sciarrino,
 Quantum walks of two correlated photons in a 2D synthetic lattice, npj Quantum Inf. {\bf 8}, 34 (2022).\\
26. H.B. Rosenstock,
Random Walks on Lattices with Traps,
J. Math. Phys. {\bf 11}, 487 (1970).\\
27.G.H. Weiss, Asymptotic form for random walk survival probabilities on three-dimensional lattices with traps, Proc. Natl. Am. Soc. {\bf 77}, 4391 (1980).\\
28. B. Ya. Balagurov and V.G. Vaks, Random walks of a particle on lattices with traps,  Sov. Phys.JETP, {\bf 38},  968 (1974).\\
29. G.H. Weiss and S. Havlin, Trapping of random walks on the line, J. Stat. Phys. {\bf 37}, 17 (1984).\\
30. S. Havlin, M. Dishon, J.E. Kiefer, and G.H. Weiss,
Trapping of Random Walks in Two and Three Dimensions,
Phys. Rev. Lett. {\bf 53}, 407 (1984).\\
31.O.E. Percus, Phase Transition in One-Dimensional Random Walk with Partially Reflecting Boundaries, Adv. Appl. Prob.{\bf 17}, 594 (1985).\\
32. E. Bach, S. Coppersmith, M.P. Goldschen, R. Joynt, and J. Watrous,
One-dimensional quantum walks with absorbing boundaries,
J. Comp. System Sci. {\bf 69}, 562 (2004).\\
33. M. G\"on\"ulol, E. Aydiner, and O.E. M\"ustecaploglu, Decoherence in two-dimensional quantum random walks with traps,
Phys. Rev. A {\bf 80}, 022336 (2009).\\
34. M. G\"on\"ulol, E. Aydiner, Y. Shikano, and O.E. M\"ustecaploglu,
Survival probability in a one-dimensional quantum
walk on a trapped lattice, New J. Phys. {\bf 13},  033037 (2011).\\
35. P. L. Krapivsky, J. M. Luck, and K. Mallick, Survival of classical and quantum particles in the presence
of traps, J. Stat. Phys. {\bf 154} 1430 (2014).\\
36. L. Regnier, M. Dolgushev, S. Redner, and O. Benichou, 
 Universal exploration dynamics of random walks,
Nature Commun. {\bf 14},  618 (2023).\\
37. M.S. Rudner and L.S. Levitov,
Topological Transition in a Non-Hermitian Quantum Walk, 
Phys. Rev. Lett. {\bf 102}, 065703 (2009).\\
38.  A. Regensburger, C. Bersch, B. Hinrichs, G. Onishchukov, A. Schreiber, C. Silberhorn, and U. Peschel, Photon propagation in a discrete fiber network: an interplay of coherence and losses, Phys. Rev. Lett. {\bf 107}, 233902 (2011).\\
39. M. Wimmer, A. Regensburger, M.-A. Miri, C. Bersch, D.N. Christodoulides, and U. Peschel, Observation of optical solitons in PT-symmetric lattices, Nat. Commun. {\bf 6}, 7782 (2015).\\
40.  S. Weidemann, M. Kremer, S. Longhi, and A. Szameit,
 Coexistence of dynamical delocalization and spectral localization through stochastic dissipation,
Nature Photon. {\bf 15}, 576 (2021).\\
41. S. Wang, C. Qin, W. Liu, B. Wang, F. Zhou, H. Ye, L. Zhao, J. Dong, X. Zhang, S. Longhi, and P. Lu, High-order dynamic localization and tunable temporal cloaking in ac-electric-field driven synthetic lattices, Nature Commun. {\bf 13}, 7653 (2022).\\
42. S. Longhi, Robust Anderson transition in non-Hermitian photonic quasicrystals, Opt. Lett.  (in press, 2024); preprint available in opticaopen.\\
43. R. A. Carmona and S. A. Molchanov, Stationary parabolic Anderson model and intermittency, Probab. Theory Relat. Fields {\bf 102}, 433 (1995).

\end{document}